\pdfoutput=1
\documentclass[12pt]{article}
\usepackage[margin=1in]{geometry}
\usepackage{amsmath}
\usepackage{titling}
\usepackage{blindtext}
\usepackage{pgfplots}
\usepackage{natbib}
\usepackage{comment}
\usepackage[colorlinks=true,linkcolor=blue,allcolors=blue]{hyperref}
\usepackage{url}
\usepackage{setspace}
\usepackage{authblk}
\begin{document}

\title{Simulating Solar Coronal Mass Ejections constrained by Observations of their Speed and Poloidal flux}


\author[1]{T. Singh}
\affil[1]{Department of Space Science, The University of Alabama in Huntsville, AL 35805, USA}

\author[2]{M. S. Yalim}
\affil[2]{Center for Space Plasma and Aeronomic Research, The University of Alabama in Huntsville, AL 35805, USA}

\author[1,2]{N. V. Pogorelov}

\author[3]{N. Gopalswamy}
\affil{NASA/Goddard Space Flight Center, Greenbelt, MD 20771}

\setcounter{Maxaffil}{0}
\renewcommand\Affilfont{\itshape\small}
\date{}    
\begin{titlingpage}
    \maketitle
\begin{abstract}

We demonstrate how the parameters of a Gibson-Low flux-rope-based coronal mass ejection (CME) can be constrained using remote observations. Our Multi Scale Fluid-Kinetic Simulation Suite (MS-FLUKSS) has been used to simulate the propagation of a CME in a data driven solar corona background computed using the photospheric magnetogram data. We constrain the CME model parameters using the observations of such key CME properties as its speed, orientation, and poloidal flux. The speed and orientation are estimated using multi-viewpoint white-light coronagraph images. The reconnected magnetic flux in the area covered by the post eruption arcade is used to estimate the poloidal flux in the CME flux rope. We simulate the partial halo CME on 7 March 2011 to demonstrate the efficiency of our approach. This CME erupted with the speed of 812 km/s and its poloidal flux, as estimated from source active region data, was $4.9\times10^{21}\, \textup{Mx}$. Using our approach, we were able to simulate this CME with the speed 840 km/s and the poloidal flux of $5.1\times10^{21}\, \textup{Mx}$, in remarkable agreement with the observations.


\end{abstract}
\end{titlingpage}



\section{Introduction} \label{sec:intro}


Predictions of the coronal mass ejection (CME) arrival times and their properties at 1 AU are of extreme importance because CMEs are the major drivers of extreme space weather events. A CME can dramatically change the near-Earth plasma conditions resulting in adverse impacts on space assets. A lot of energy can be deposited in the Earth's upper atmospheric layers potentially resulting in communication losses with satellites, which have become integral parts of our technologically advanced society. 

MHD simulations are very promising for CME forecasting due to the advancement in parallel computation techniques and better CME models. Many observed CMEs have been successfully simulated using MHD models where several properties of simulated and observed CMEs were matched with reasonable accuracy ~\citep[][and references therein]{Manchester04a, Jin17b, Si18, Scolini18}.  

Current CME models used in MHD simulations can be broadly divided into two categories: (1) over-pressured plasmoid models, such as blob model~\citep[e.g., see,][]{Chane05,OP99a} and (2) flux-rope-based models, such as the Gibson--Low (GL) \citep{GL98} model, the Titov--Demoulin model \citep{TD99}, and their variations. Since the magnetic flux rope of a CME is a major contributor to its propagation and impact with Earth's magnetosphere, flux-rope-based models are clearly more realistic and promising for space weather prediction. Furthermore, there is strong observational evidence that all CMEs reaching the interplanetary medium have flux-rope structure \citep{Gopalswamy13a}. In particular, flux rope models can be readily used to calculate the $z$-component of magnetic field at 1 AU, which determines the geoeffectiveness of magnetic storms. In addition, it is important to develop such models at distances close to the solar surface because high energy solar energetic particles are typically accelerated at distances below 5 $R_{\odot}$ \citep{Gopalswamy13b}.\\
Some previous works ~\citep[e.g.,][]{Si18, Jin17a} were successful in matching the CME speeds and orientations, but no attention was paid to matching the poloidal flux of CME flux ropes. To address this shortcoming, we propose a method to find the input parameters for the Gibson--Low flux-rope-based model by using the observations of CME speed, orientation, and magnetic poloidal flux, and matching them to those in the simulated CME. We build on a new method developed by \citet{Gopalswamy18}, and called Flux Rope from Eruption Data (FRED), to find the poloidal flux of an erupted CME. Then, the parametric study similar to that performed by \citet{Si18} is used to constrain the GL parameters.

Section \ref{obs} describes how the CME parameters are derived from observations. Section \ref{models} describes the background solar wind and CME models used in this study. Section \ref{constrain} addresses the proposed method for deriving the flux rope parameters that match observations. Finally, in Section \ref{conc}, we apply this approach to an observed CME and present our conclusions.     

\section{CME observations} \label{obs}
In this study, we pay attention to three major physical properties of CMEs: their speed, orientation angle, and flux rope poloidal flux. These properties can be estimated by using the Solar Terrestrial Relations Observatory (\textit{STEREO}) \textit{A \& B} coronagraphs, Solar and Heliospheric Observatory (\textit{SOHO}) coronagraphs, Solar Dynamics Observatory-Atmospheric Imaging Assembly (\textit{SDO-AIA}) data in 193 \r{A}, and \textit{SDO}-Helioseismic and Magnetic Imager (\textit{HMI}) magnetograms.  If we approximate a CME as a structure with two conical legs and a curved front, the CME orientation angle can be defined as the angle the plane containing CME legs makes with the local longitude line. The orientation angle is represented by $\gamma$ in Fig. 3 of \cite{THV06}. CME rotation can change this angle.\\

We calculate the true speed and the orientation of the CME with respect to the local longitude line using the Graduate Cylindrical Shell (GCS) fitting method \citep{THV06, Thernisien11}. GCS fitting is a visual tool in which three viewpoints of a CME from \textit{STEREO A \& B} and \textit{SOHO} coronagraphs are used to fit its structure with conical legs and a curved surface embracing this CME, as seen from all three viewpoints. By fitting the CME properties with the GCS model for a series of observational time frames, we can obtain the CME height-time dependence and calculate its speed by linear fitting. The orientation angle determined with this method is used when the derived CME is inserted into the background solar wind.\\
The poloidal flux of a CME can be found by the method described by \citet{Gopalswamy17}. They find the reconnected flux of a flux rope by calculating the unsigned magnetic flux in the area covered by the post eruption arcade when a CME leaves the solar surface. The reconnected flux has been shown to be approximately equal to the poloidal flux of a CME by \citet{Qiu07}. We can find the poloidal flux of an erupted CME from SDO HMI line-of-sight magnetograms and EUV 193 \r{A} images of the source active region. This method works best when the source active region of a CME lies roughly within $30^\textup{o}$ from the solar disk center in both latitude and longitude. This is because the accuracy of magnetograms decreases as one moves away from the solar disk center. Since the majority of Earth-directed CMEs originate near the disk center, this method is extremely useful for space weather predictions.   

\section{Physical models} \label{models}
\subsection{Corona model}

We use our global MHD model of the solar corona \citep{YPL17} implemented in the Multi-Scale Fluid-Kinetic Simulation Suite (MS-FLUKSS, \cite{Pogorelov14}). This model is designed to be driven by a variety of observational solar magnetogram data. In this study, we use SDO HMI synoptic radial magnetogram data. We solve the set of ideal MHD equations in a frame of reference corotating with the Sun and use the volumetric heating source terms \citep{Nakamizo09} to model solar wind acceleration. We calculate the initial solution for the magnetic field using the Potential Field Source Surface (PFSS) model \citep{TVH11}. For the rest of plasma parameters, we compute the initial solution from Parker's isothermal solar wind model \citep{Parker58}. The boundary conditions are described later in Section \ref{conc}.

\subsection{The Gibson-Low flux rope model}

The magnetostatic solution to the Gibson-Low flux rope problem is found by balancing the magnetic, pressure gradient, and gravitational forces, i.e.,  $(\nabla\times\mathbf{B})\times\mathbf{B}-\nabla p-\rho\mathbf{g}=0$, where $\mathbf{B}$, $p$, $\rho$, and $\mathbf{g}$ are the magnetic field, plasma pressure, density, and the gravitational acceleration, respectively. The condition of solenoidal magnetic field, $\nabla\cdot\mathbf{B}=0$,  is also taken into account. This gives us an analytical solution for $\mathbf{B}$  in the form of a spherical torus which is further stretched using the transformation $r\rightarrow r-a$ in spherical coordinates. Here $r$ is the radial coordinate and $a$ is the stretching parameter. This creates a magnetic field line distribution of a tear-drop shape. The analytical solution for a GL flux rope involves four parameters:
\begin{enumerate}
\item Flux rope radius ($r_0$): This is the radius of an initial GL spherical torus before stretching.
\item Flux rope height ($r_1$): This is the height of the center of the introduced spherical torus with respect to the center of the Sun before stretching.
\item Flux rope stretching parameter ($a$): This is the amount by which each part of the spherical torus is stretched towards the center of the Sun.
\item Flux rope field strength ($a_1$): This is a free parameter that controls the field strength in the flux rope. Plasma pressure inside the rope is proportional to $a_1^2$ due to the condition of pressure balance assumed in this solution.   
\end{enumerate}

\begin{figure}

\center
\begin{tabular}{c c}  

\includegraphics[scale=0.1,angle=0,width=6cm,keepaspectratio]{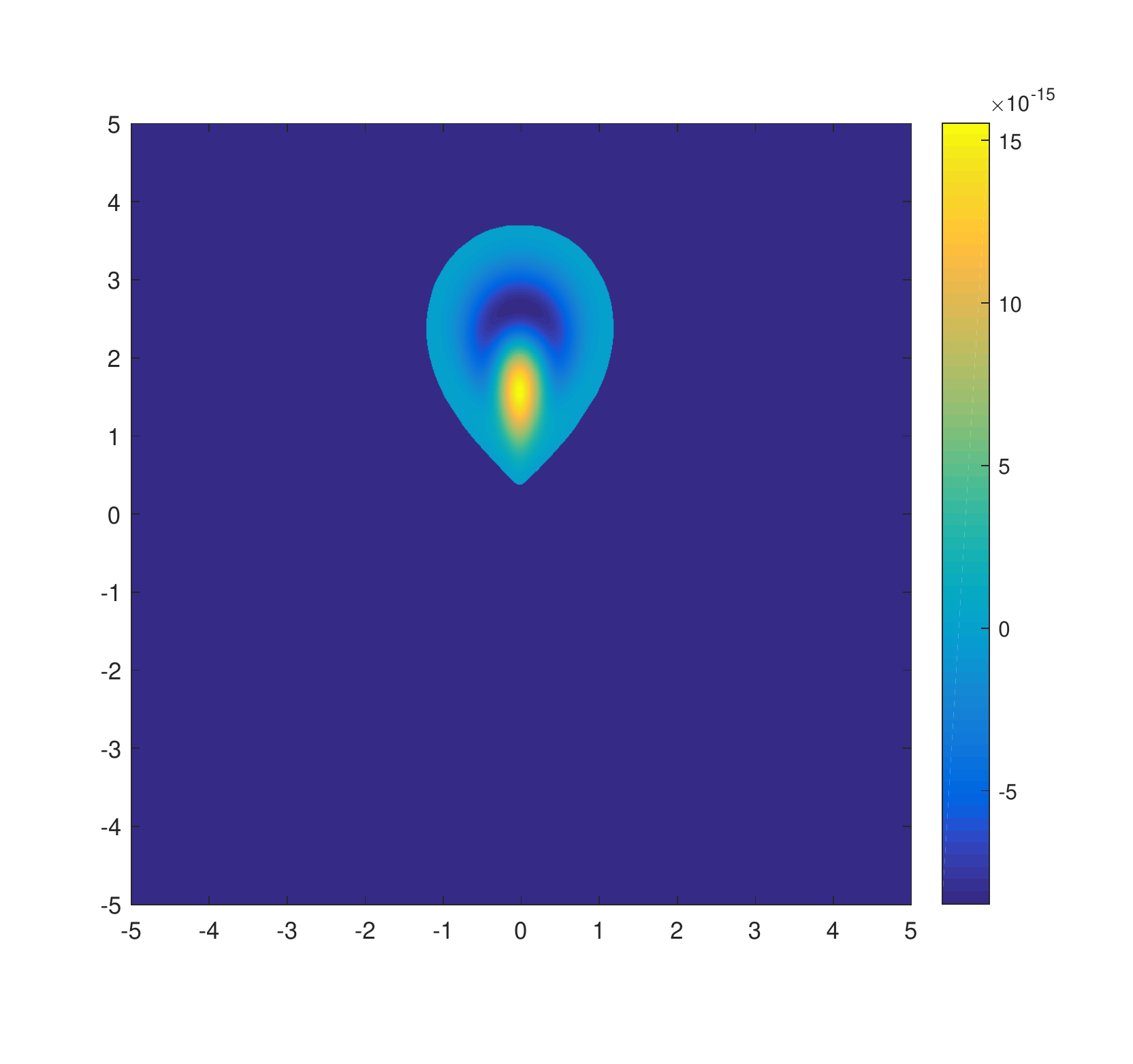}
\includegraphics[scale=0.1,angle=0,width=6cm,keepaspectratio]{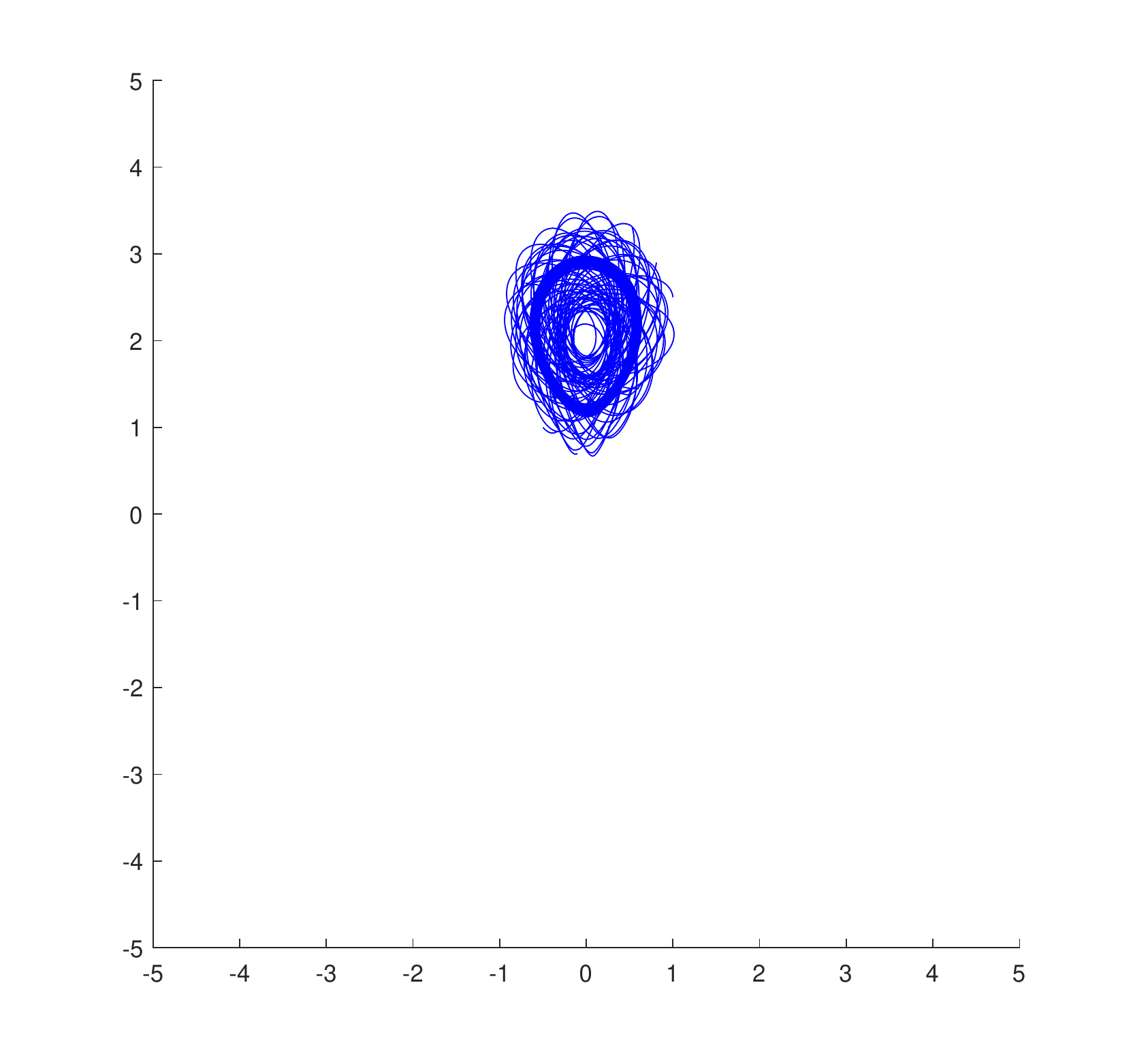} 

\end{tabular}

\caption{Density ($g/cm^3$)(\textit{left}) and magnetic field lines (\textit{right}) in a stretched GL torus. All horizontal and vertical axes are in $R_{\odot}$. We used $r_0=1.67$, $r_1=3.03$, $a=1.01$ and $a_1=0.23$ in these figures. The origin is located at the solar center.}
\label{GL0}
\end{figure}

Figure \ref{GL0} shows density and magnetic field lines of GL torus after stretching. We notice that the density distribution in a stretched flux rope resembles widely accepted 3-part structure of a CME: its bright front, dark cavity, and bright core.

A GL flux rope is inserted into the solar wind background by superposition such that the pointed end of the tear drop shaped flux rope is under the solar surface. This results in anchoring of two flux rope legs on the Sun (See Figure \ref{expansion}). When the solution is evolved with time, the flux rope is no longer in equilibrium and erupts as a CME. The analytical solution of a GL model can have negative density and pressure in some regions of the flux rope. To avoid this problem, we follow \citet{Manchester04b} and make sure that total density and pressure do not fall below 25\% of the density and pressure in the original background. 

\section{Constraining GL parameters} \label{constrain}

Previous works ~\citep[e.g., ][]{Si18, Jin17a} show that the speed of a simulated CME depends on the GL parameters. In particular, \citet{Si18} demonstrated that it depends on $a_1r_0^4$ linearly. The dependence on $r_1$ is linear for $r_1 < 2.6 R_{\odot}$ and independent of $r_1$  for $r_1 \geq 2.6 R_{\odot}$. They also found that the speed of a CME is inversely  correlated with the average thermal pressure of the background solar wind in the initial direction of CME propagation. These results were then used to find an empirical relation for the simulated CME speed:
\begin{equation}
V_\mathrm{CME}=\begin{cases}
(c_1a_1r_0^4+c_2)\cdot(c_3P_{avg}+c_4)\cdot(c_5r_1+c_6) &r_1<2.6 \\
(c_1a_1r_0^4+c_2)\cdot(c_3P_{avg}+c_4) &r_1\geq 2.6
\end{cases}
\label{eq1}
\end{equation}
where $c_1,\dotsc ,c_6$ were found using the non-linear multi-variable regression. Since CMEs propagate through the low-$\beta$ corona, it makes more sense to use the total pressure rather than just the thermal pressure. We find that the total pressure is inversely correlated with the CME speed with correlation coefficient of -0.97. We have redone our analysis using the new definition and updated the fitting constants accordingly. They are found to be \{3.56, 5.39, -0.06, 24.63, 2.57, 5.56\} for $r < 2.6$ and \{12.88, 19.52, -0.20, 52.97\} for $r \geq 2.6$. 

  
 
  


We can simulate a CME that matches the observed speed and poloidal flux as follows. Firstly, we iterate over a set of reasonable GL parameters and find the corresponding speed and poloidal flux of the simulated CME. This speed is given by Eq. (\ref{eq1}) and  the poloidal flux is found using the GL analytical solution. After that, there only remains to choose the set of parameters which matches the observations the best. Although this method is rather simple, it may significantly improve the quality of space weather predictions. This is because we are constraining not only the speed of the CME, which naturally affects the CME arrival time at 1 AU, but also the poloidal flux and orientation of CME flux rope, which affect $B_z$ at 1 AU. The sign and magnitude of $B_z$ are very important quantities that control the geoeffectiveness of a CME. It has been found that CMEs with negative (southward) $B_z$ flux ropes are more geoeffective due to the favorable coupling with the Earth's magnetosphere \citep{Burton75}. Therefore, an attempt to match poloidal flux in CME models will be helpful in predicting $B_z$ values at 1 AU, and therefore, the CME geoeffectiveness.    

\section{Results and Conclusions} \label{conc}

We now apply our approach to a CME that erupted on March 7, 2011 at 14:00 UT. It erupted from active region AR-11166 which was then at ~N11E20 on the solar disk. First, we generate an MHD corona background solution driven by SDO HMI synoptic radial magnetogram for CR 2107 at the inner boundary located at the lower corona just above the transition region \citep{Nakamizo09}. This solution is obtained by relaxing the initial PFSS magnetic field distribution to a steady state using our solar corona model. We used the TVD, finite-volume Rusanov scheme~\citep{KPS01} to compute the numerical fluxes and the forward Euler scheme for time integration. In order to satisfy the solenoidal constraint, we applied Powell's source term method~\citep{Powell99}. The size of our computational domain is 1.03$R_{\odot}\leq$ r $\leq 30R_{\odot}$, $0\leq\phi\leq 2\pi$, $0\leq\theta\leq\pi$. We use the grid of 180$\times$240$\times$120 in r, $\phi$ and $\theta$ directions, respectively. We perform all simulations in the frame corotating with the Sun. MS-FLUKSS ensures an efficient parallel implementation of our numerical methods. At the inner boundary of the computational domain, which is located at the lower corona, we specify the radial magnetic field derived from the HMI line-of-sight magnetogram data and the differential rotation~\citep{KHH93a} and meridional flow~\citep{KHH93b} formulae for determining the horizontal velocity components at the ghost cell centers. We kept density and temperature constant as $n=1.5 \times 10^{8} \,\textrm{cm}^{-3}$ and $T=1.3 \times 10^{6}$ K, respectively. The radial velocity component is imposed to be zero at the boundary surface. The transverse magnetic field components are extrapolated from the domain into the ghost cells below the inner radial boundary. At the outer boundary of the domain, which is located beyond the critical point, the plasma flow is superfast magnetosonic, so no boundary conditions are required.

The observed CME shows up in the STEREO A \& B Cor2 coronagraph FOV at 14:54 UT. We use the GCS model to calculate speed of CME as 812 km/s. The orientation of the erupted CME is estimated to be $5^\circ$ clockwise from local longitude line. The CME was found to be traveling radially in direction of $18^\circ$ North and $24^\circ$ East. It is important to note that these values are different from those one might get by using only the pre-eruption neutral line (NL) location and orientation, which is done, for example, in the Eruptive Event Generator Gibson Low (EEGGL) approach made available at the Community-Coordinated Modeling Center (CCMC) \citep[e.g.,][]{Jin17a, Borovikov17, Borovikov18}. The EEGGL approach fixes the GL parameters $r_1$ and $a$ as 1.8 and 0.6, respectively. The size parameter $r_0$ is estimated from the source active region NL length using a scaling factor. Finally, the magnetic strength parameter $a_1$ is constrained using observed CME speed via a parametric study. The EEGGL model also assumes the orientation angle and direction of the CME at initial height to be the same as NL. Our current approach treats parameters  $r_1$ and $a$ as variables and is more flexible. Also, our approach uses the GCS model that takes into account the deflection and rotation, which a CME may experience by the time it reaches the coronagraph FOV. \citet{Kay15} have found that CMEs undergo maximum deflection in low corona ($<\,5\,R_{\odot}$), with the primary deflection occurring below 3 $R_{\odot}$ for strong background fields. Similarly, CMEs can experience rotation in their orientation very low in the corona \citep{Manchester17}. For example, \citet{Thompson12} reported a CME that rotated $115^\circ$ within a distance of 1.5 $R_{\odot}$ from the solar surface. When the GL flux rope model is introduced into the corona, its top edge is already more than 1 $R_{\odot}$ from the solar surface. By this time, a CME's orientation and direction can be significantly different from the one found using NL data. Therefore, direction and orientation inputs from GCS model are more reliable than from NL inputs.\\

\begin{figure}[ht]
\center
\includegraphics[scale=0.4,angle=0,width=15cm,height=15cm,keepaspectratio]{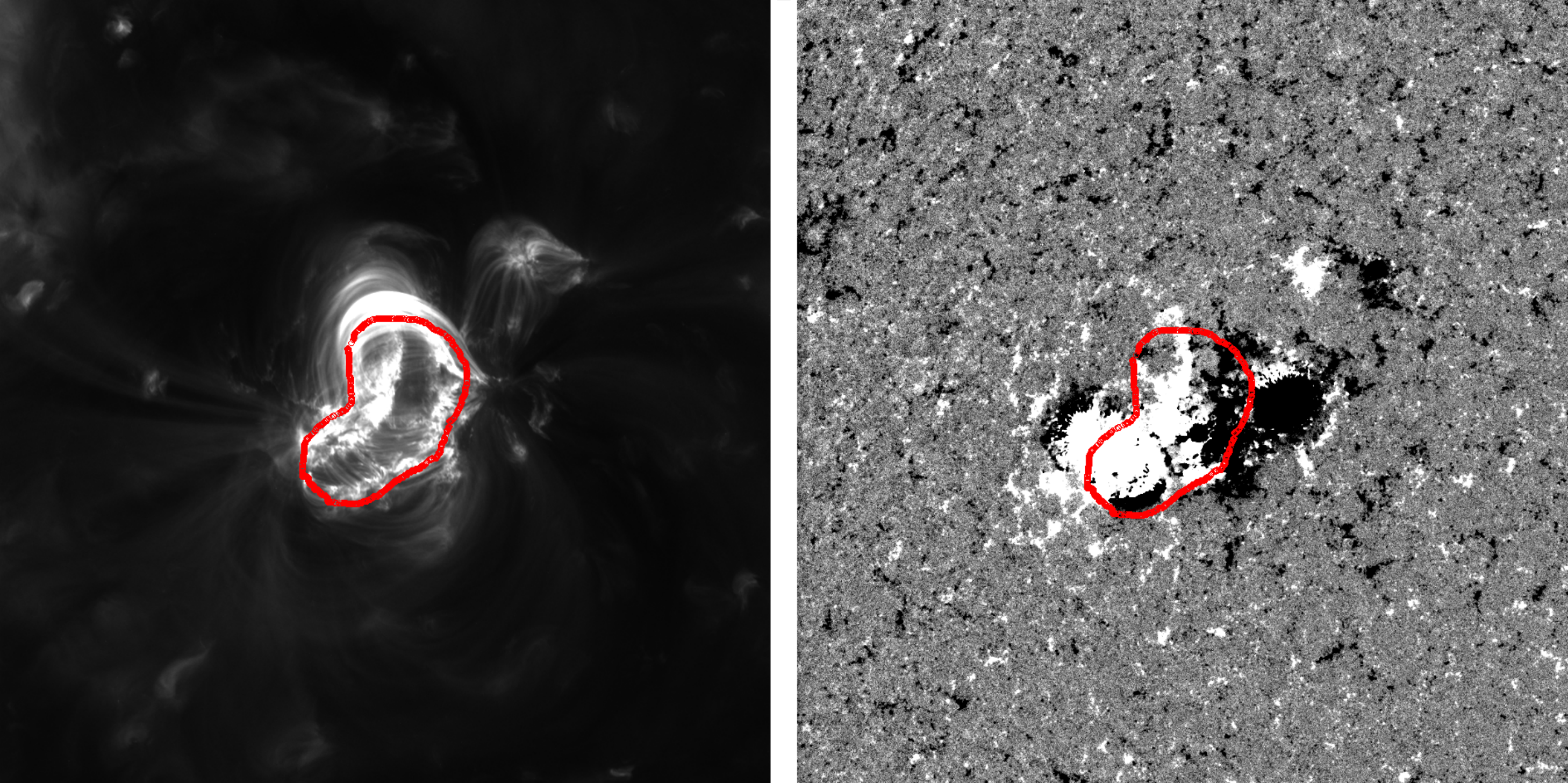} 
\caption{$(left)$ Post eruption arcade as seen in SDO AIA 193 imager at 16:21 UT. The area between the footpoints is enclosed by red points. $(right)$ The same area is used in pre-eruption SDO HMI line-of-sight magnetogram at 13:00 UT to find reconnected flux. The box size in both frames is 400 Mm.}
\label{fred}
\end{figure}

We use CME brightness in Cor2 coronagraphs of STEREO A \& B to estimate the mass of the CME as $3.9\times10^{12}$ kg, which is not too different from the mass listed in the SOHO/LASCO CME catalog (\url{https://cdaw.gsfc.nasa.gov/CME_list}). We use the ``Flux Rope from Eruption Data'' approach to estimate the poloidal flux of the CME to be $4.9\times10^{21}\, \textup{Mx}$. Figure \ref{fred} shows the post-eruption arcade area after the eruption in the source active region and the corresponding pre-eruption magnetogram used to estimate the poloidal flux. From the solar corona background, we estimate the average pressure of the solar wind to be 67.1 mdyn/cm$^2$.

\begin{figure}

\center
\begin{tabular}{c c c}  

\includegraphics[scale=0.1,angle=0,width=5.cm,keepaspectratio]{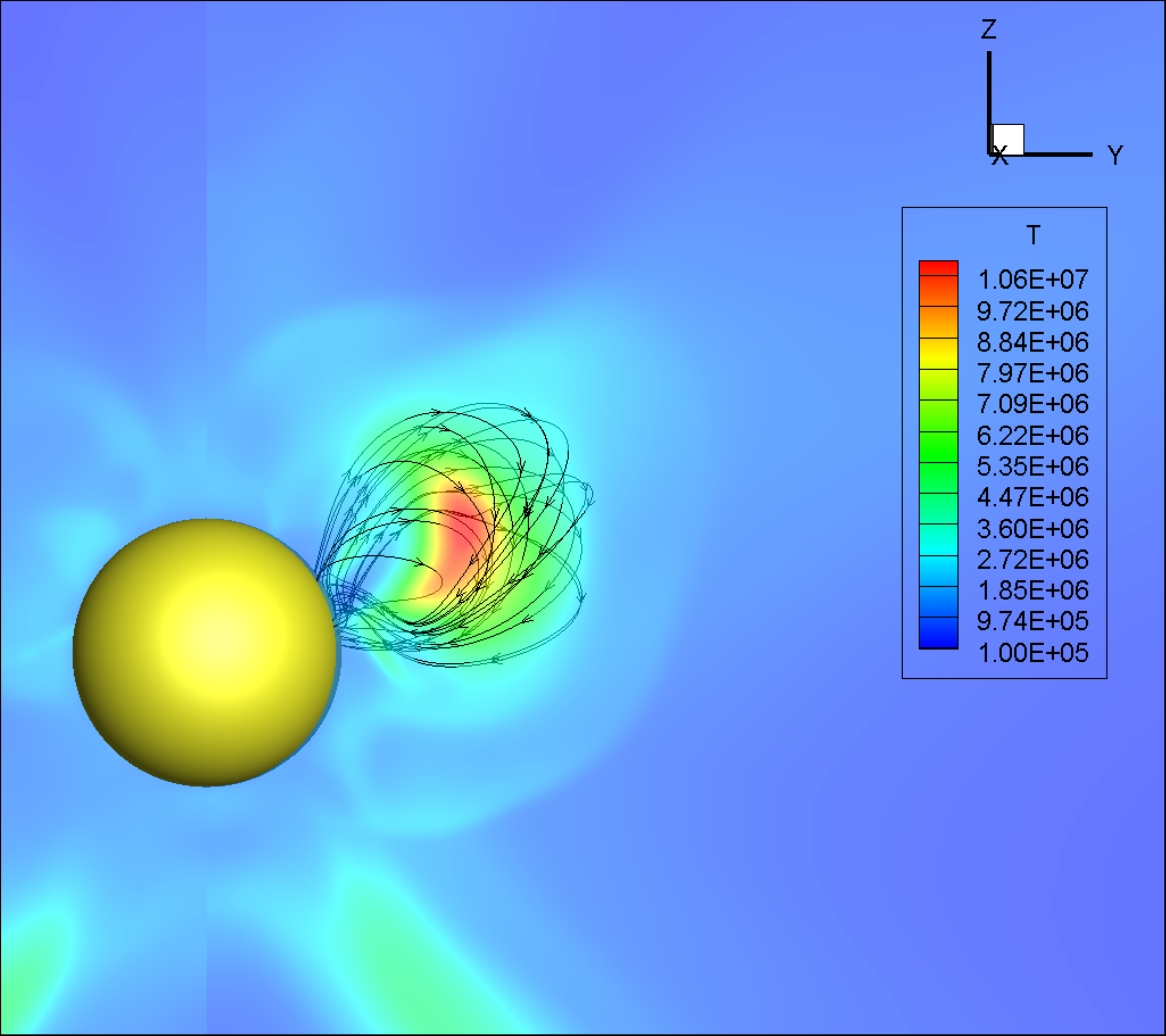}
\includegraphics[scale=0.1,angle=0,width=5.cm,keepaspectratio]{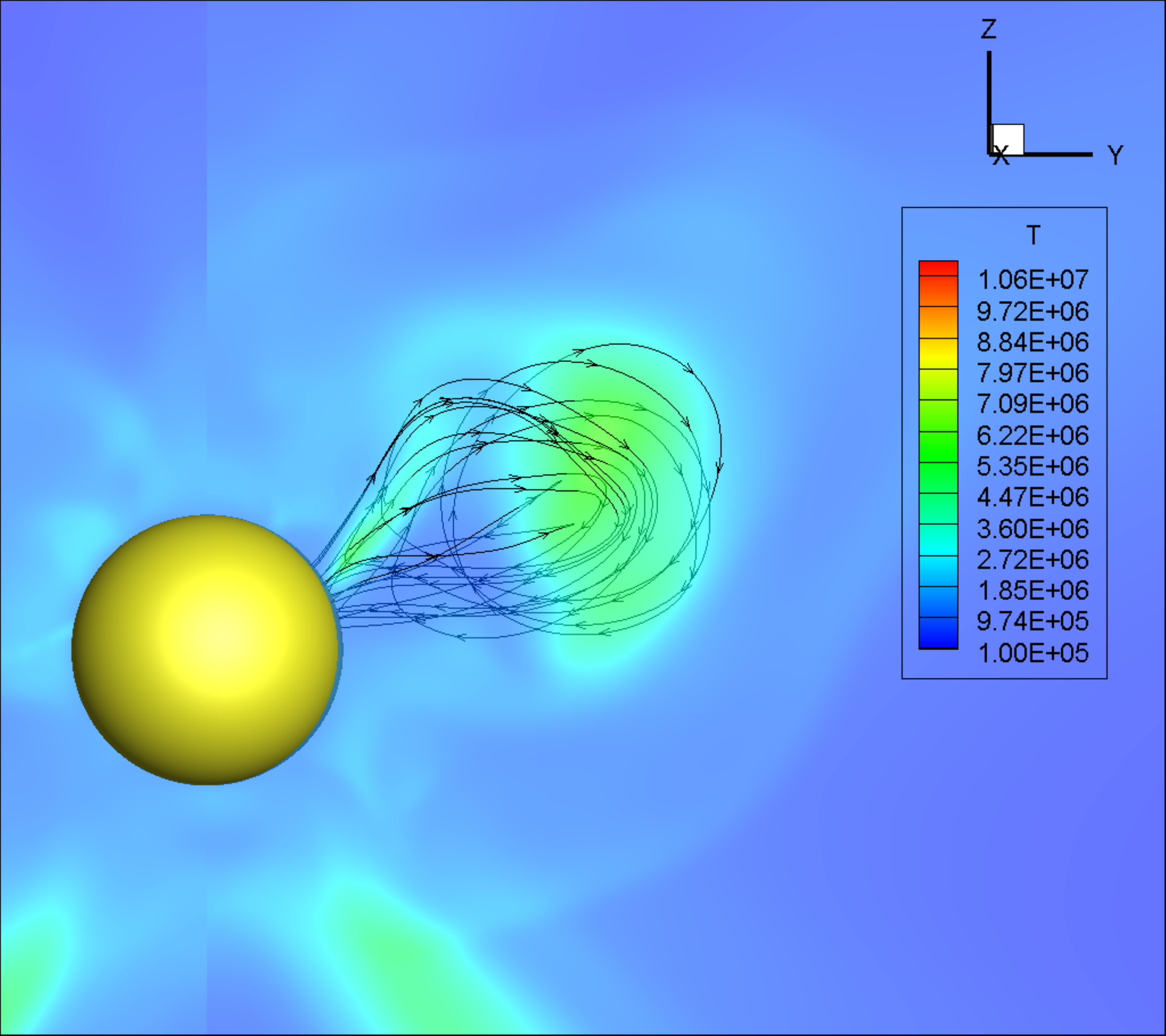} 
\includegraphics[scale=0.1,angle=0,width=5.cm,keepaspectratio]{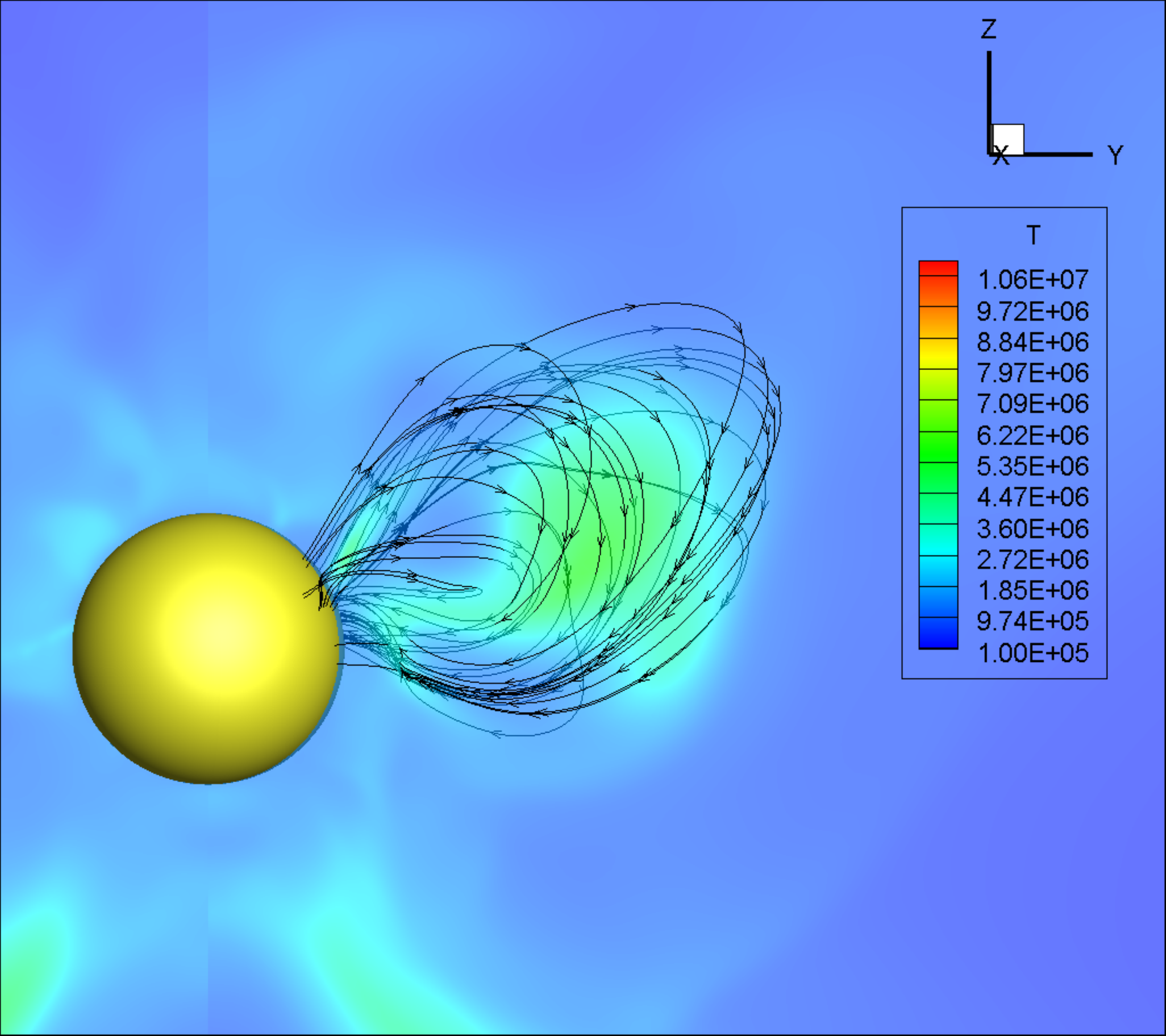} 
\end{tabular}

\caption{From \textit{left} to \textit{right}: Flux rope of the simulated CME after 36, 69 and 101 minutes after initial insertion. We also show a translucent slice in the plane of CME propagation, perpendicular to the ecliptic plane, with temperature (K) contours. The yellow sphere represents the Sun. The frame size is 9.2 $R_{\odot}$ $\times$ 7.9 $R_{\odot}$.}
\label{expansion}
\end{figure}

\begin{figure}[ht]
\center
\includegraphics[scale=0.1,angle=270,width=12cm,keepaspectratio]{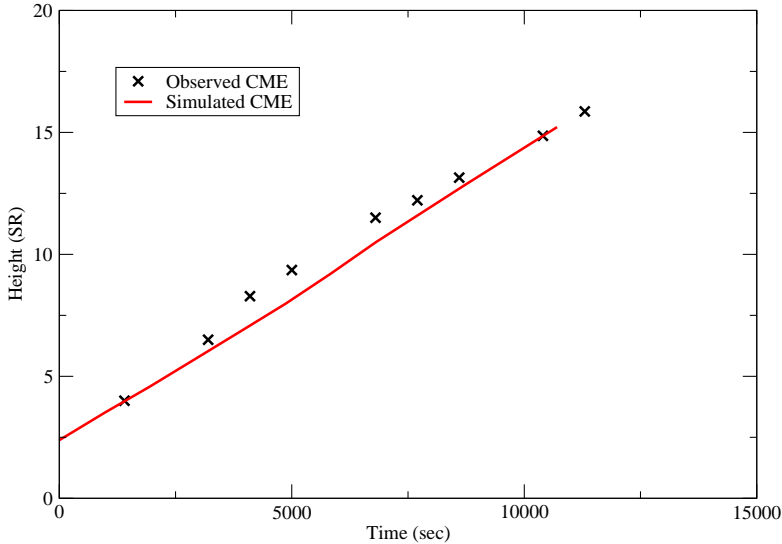} 
\caption{Comparison of height vs time evolution between the observed and simulated CMEs. Heights of observed CME at each time were estimated by GCS fitting.}
\label{ht}
\end{figure}

Now, we apply the approach described in Section \ref{constrain} to find the GL parameters. We iterate $a_1$ from 0.1 to 3.2, $r_0$ from 0.1 to 3.1, $r_1$ from 1.0 to 3.0, and $a$ from 0.1 to 2.5, all with the steps equal to 0.1. We look for GL parameter sets that give us the poloidal flux and speed within a 10\% error margin from their observed values. Clearly, there is no unique set of such parameters. We list a few of them in Table \ref{tab2}. We also include the poloidal flux introduced related to these sets, as well as their estimated speeds defined by Equation \ref{eq1}. We randomly choose $r_0 = 1.0$, $r_1 = 2.4$, $a = 1.1$,  and $a_1 = 0.2$ from our list of acceptable GL parameters. As seen from Table \ref{tab2}, these parameters should end up simulating a CME with speed  782 km/s and poloidal flux of $5.1\times10^{21}$ Mx. We inserted this GL flux rope into the solar wind background in direction $18^\circ$ North and $24^\circ$ East and orientation $5^\circ$ clockwise as was estimated by GCS model. These GL parameters give us the simulated CME with the speed of 840 km/s. Figure \ref{expansion} shows the time evolution of the simulated CME. Figure \ref{ht} compares the height-time evolution of the simulated CME with observed values found using GCS fitting. The agreement between the observed and simulated CMEs, as they propagate from the Sun, is clearly seen.  

\begin{table}[]
\centering
\begin{tabular}{|c|c|c|c|c|c|c|}
\hline
$a_1$ ($\textup{Gauss}/R_{\odot}^2$)  & $r_0$ ($R_{\odot}$)  & $r_1$ ($R_{\odot}$) & $a$ ($R_{\odot}$)  & Pol\_flux (Mx) & Mass (g)     & Speed (km/s) \\ \hline
0.3 & 0.9 & 2.4 & 1.2 & 4.64E+21  & 5.35E+15 & 781   \\ \hline
0.3 & 0.9 & 2.5 & 1.3 & 4.64E+21  & 6.39E+15 & 798   \\ \hline
0.3 & 0.9 & 2.6 & 1.4 & 4.63E+21  & 7.43E+15 & 828   \\ \hline
0.6 & 0.8 & 2.8 & 1.7 & 5.09E+21  & 7.23E+15 & 852   \\ \hline
0.2 & 1   & 2.4 & 1.1 & 5.07E+21  & 6.78E+15 & 782   \\ \hline
0.2 & 1   & 2.4 & 1.2 & 4.62E+21  & 1.73E+15 & 782   \\ \hline
0.2 & 1   & 2.5 & 1.3 & 4.62E+21  & 2.53E+15 & 799   \\ \hline
0.2 & 1   & 2.6 & 1.4 & 4.61E+21  & 3.34E+15 & 830   \\ \hline
0.1 & 1.2 & 2.5 & 1.2 & 5.07E+21  & 5.66E+14 & 803   \\ \hline
0.1 & 1.2 & 2.6 & 1.3 & 5.06E+21  & 1.48E+15 & 834   \\ \hline
0.1 & 1.2 & 2.7 & 1.4 & 5.05E+21  & 2.39E+15 & 834   \\ \hline
0.1 & 1.2 & 2.8 & 1.5 & 5.05E+21  & 3.31E+15 & 834   \\ \hline
0.1 & 1.2 & 2.9 & 1.6 & 5.04E+21  & 4.24E+15 & 834   \\ \hline
\end{tabular}
\caption{Parameters used in the expression of $V_{CME}$.}
\label{tab2}
\end{table}

This example shows the power of the proposed method when applied to simulations of CMEs with the speed, poloidal flux, orientation, and direction derived from observations. These CME properties have direct impact on its geo-effectiveness. The CME can be propagated further to 1 AU and its properties such as  arrival time and $B_z$ can be measured. The proposed approach can be further improved by using the observed CME mass. In a recent paper of \citet{Howard18}, it was found that there is no pileup of mass in front of propagating CMEs up to height of 30 $R_{\odot}$. This means that the mass of a CME observed in the coronagraph, when it has completely emerged from behind the coronagraph occulter, can be matched with the mass introduced in the GL model. Typically, this occurs when the CME front has reached a height of 7 $R_{\odot}$. Another observable that can be constrained is the angular width of a CME. In \citet{Si18}, the stretching parameter $a$ was fixed as $a=r_1/3$. Since this parameter controls the shape and angular width of initial GL flux rope, further study needs to be done on how this parameter impacts the expansion of the CME.      However, even the addition of poloidal flux as a constraining parameter makes the GL model closer to reality.


TS acknowledges the graduate student support from NASA Earth and Space Science Fellowship. The authors acknowledge the support from the UAH IIDR grant 733033. This work is also supported by the Parker Solar Observatory contract with the Smithsonian Astrophysical Observatory through subcontract SV4-84017. We also acknowledge NSF PRAC award OAC-1811176 and related computer resources from the Blue Waters sustained-petascale computing project. Supercomputer allocations were also provided on SGI Pleiades by NASA High-End Computing Program award SMD-16-7570 and on Stampede2 by NSF XSEDE project MCA07S033. NG was supported in part by NASA's LWS TR\&T program.

This work utilizes data from \textit{SOHO} which is a project of international cooperation between ESA and NASA. The HMI data have been used courtesy of NASA/\textit{SDO} and HMI science teams. The \textit{STEREO}/SECCHI data used here were produced by an international consortium of the Naval Research Laboratory (USA), Lockheed Martin Solar and Astrophysics Lab (USA), NASA Goddard Space Flight Center (USA), Rutherford Appleton Laboratory (UK), University of Birmingham (UK), Max-Planck-Institut for Solar System Research (Germany), Centre Spatiale de Li\`ege (Belgium), Institut d'Optique Th\'eorique et Appliqu\'ee (France), and Institut d'Astrophysique Spatiale (France). This work uses SOHO CME catalog which is generated and maintained at the CDAW Data Center by NASA and The Catholic University of America in cooperation with the Naval Research Laboratory.




\end{document}